\documentclass[fleqn,usenatbib]{mnras}
\usepackage{newtxtext,newtxmath}
\usepackage[T1]{fontenc}
\usepackage{ae,aecompl}

\usepackage{etoolbox}
\makeatletter
\patchcmd\@combinedblfloats{\box\@outputbox}{\unvbox\@outputbox}{}{\errmessage{\noexpand patch failed}}
\makeatother

\usepackage{graphicx}	% Including figure files
\usepackage{amsmath}	% Advanced maths commands
\usepackage{amssymb}	% Extra maths symbols
\usepackage{soul}
\usepackage{newtxtext}
\usepackage[english]{babel}
\usepackage{booktabs}
\usepackage{graphicx,color}
\usepackage{epstopdf}
\newcommand{\Mpch}{$h^{-1}\,\mbox{Mpc}$}

\newcommand{\xiis}{$\xi(s_\perp,s_\parallel)$}
\newcommand{\Om}{$\Omega_{\rm M}$}
\newcommand{\Ol}{$\Omega_{\Lambda}$}

\newcommand{\dustp}{{\small DUSTGRAIN}-\emph{pathfinder}~}

\title[Clustering in MG with massive neutrinos]{Clustering and redshift-space distortions in modified gravity models with massive neutrinos}
\author[Garc\'ia-Farieta J.E. et al. 2019]{\parbox{\textwidth}{Jorge Enrique Garc\'ia-Farieta$^{1,2}$\thanks{E-mail: \href{joegarciafa@unal.edu.co}{joegarciafa@unal.edu.co}}, Federico Marulli$^{2,3,4}$, Alfonso Veropalumbo$^{5}$, Lauro Moscardini$^{2,3,4}$, Rigoberto A. Casas-Miranda$^{1}$, Carlo Giocoli$^{6,3}$ and Marco Baldi$^{2,3,4}$}\\ \\
  $^1$Departamento de F\'isica, Universidad Nacional de Colombia - Sede Bogot\'a, Av. Cra 30 No 45-03, Bogot\'a, Colombia\\
  $^2$Dipartimento di Fisica e Astronomia, Alma Mater Studiorum Universit\`{a} di Bologna, via Gobetti 93/2, I-40129 Bologna, Italy\\
  $^3$INAF - Osservatorio di Astrofisica e Scienza dello Spazio di Bologna, via Gobetti 93/3, I-40129 Bologna, Italy \\
  $^4$INFN - Sezione di Bologna, viale Berti Pichat 6/2, I-40127 Bologna, Italy \\
  $^5$Dipartimento di Fisica, Universit\`a degli Studi Roma Tre, via della Vasca Navale 84, I-00146 Roma, Italy \\
  $^6$Dipartimento di Fisica e Scienza della Terra, Universit\`a degli Studi di Ferrara, via Saragat 1, I-44122 Ferrara, Italy}
\date{Accepted XXX. Received YYY; in original form ZZZ}

\pubyear{2019}

\begin{document}
\label{firstpage}
\pagerange{\pageref{firstpage}--\pageref{lastpage}}
\maketitle

% Abstract of the paper
\begin{abstract}
Modified gravity and massive neutrino cosmologies are two of the most interesting scenarios that have been recently explored to account for possible observational deviations from the concordance $\Lambda$-cold dark matter ($\Lambda$CDM) model. In this context, we investigated the large-scale structure of the Universe by exploiting the \dustp simulations that implement, simultaneously, the effects of $f(R)$ gravity and massive neutrinos. To study the possibility of breaking the degeneracy between these two effects, we analysed the redshift-space distortions in the clustering of dark matter haloes at different redshifts. Specifically, we focused on the monopole and quadrupole of the two-point correlation function, both in real and redshift space. The deviations with respect to $\Lambda$CDM model have been quantified in terms of the linear growth rate parameter. We found that redshift-space distortions  provide a powerful probe to discriminate between $\Lambda$CDM and modified gravity models, especially at high redshifts ($z \gtrsim 1$), even in the presence of massive neutrinos.
\end{abstract}

% Select between one and six entries from the list of approved keywords.
% Don't make up new ones.
\begin{keywords}
galaxies: haloes - cosmology: theory, large-scale structure of Universe, cosmological parameters - methods: numerical, statistical
\end{keywords}

%%%%%%%%%%%%%%%%%%%%%%%%%%%%%%%%%%%%%%%%%%%%%%%%%%
%%%%%%%%%%%%%%%%% BODY OF PAPER %%%%%%%%%%%%%%%%%%
%%%%%%%%%%%%%%%%%%%%%%%%%%%%%%%%%%%%%%%%%%%%%%%%%%

\section{Introduction}\label{sec:intro}
A theory of gravity is required to describe the spatial properties and dynamics of the large-scale structure (LSS) of the Universe. The observational data collected during the last decade provided strong support to the concordance $\Lambda$CDM model, which, with only $6$ free parameters, yields a consistent description of the main properties of the LSS \citep[see e.g.][]{Tonry_SN_2003, VIPERS_2014, Subaru_lensing_2015, Planck_results_2015, Planck_params_2015, BOSS_gal_2017, CFHTLenS_params_2017, DES1_2018, Planck_params_2018, XXL_counts_2018, KiDS_params_2018, CORE_params_2018, Pan-STARRS_params_2018}. The $\Lambda$CDM model assumes General Relativity (GR) as the theory describing gravitational interactions, the standard model of particles and the Cosmological Principle, asserting that the Universe is statistically homogeneous and isotropic on large scales. In this framework, the Universe is currently dominated by the dark energy (DE), in the form of a cosmological constant, responsible for the late-time cosmic acceleration \citep{Riess_1998, Schmidt_1998ApJ, Perlmutter_1999}, and by a CDM component that drives the formation and evolution of cosmic structures.

A possible critical tension in the $\Lambda$CDM scenario consists in the discrepancy recently observed in $H_0$ and $\sigma_8$ measurements when different probes at high and low redshifts are used \citep[see][]{Planck_SZ_2015, Riess_H0_2016, Bernal_H0_2016, Planck_params_2018}. Massive neutrinos, the only (hot) dark matter (DM) candidates we actually know to exist, can affect these observables and have several cosmological implications \citep[e.g.][]{Lesgourgues_neutrinos_2006, Marulli_neutrino_2011, Costanzi_neutrino_2014, Battye_neutrino_2014, villaescusa2014, Enqvist_2015, Roncarelli2015, Zennaro_neutrinos_2018, Poulin_neutrinos_2018}. However, it has been recently shown that a strong observational degeneracy exists between some modified gravity (MG) models and the total neutrino mass \citep{Motohashi_neutrinos_2013, He_neutrinos_2013, Baldi_degeneracies_2014, Giocoli_lensing_2018A}, giving rise to an intrinsic limitation of the discriminating power of many standard cosmological statistics \citep{Peel_degeneracy_2018, Hagstotz_2018}. MG models represent one of the most viable alternatives to explain cosmic acceleration \citep[for a review see e.g.][]{Joyce_20016}. They must satisfy solar system constraints, and at the same time be consistent with the measured accelerated cosmic expansion and large-scale constraints \citep[see e.g.][and references therein]{Uzan5042, Will2014, pezzotta2017, Collett1342}.

A powerful cosmological probe to discriminate among these alternative frameworks is provided by the redshift-space galaxy clustering on different scales \citep[see e.g.][]{Arnouts_1999, Blake_2003, Percival_2007, Guzzo_2008Natur, Blake_2011, Marulli_2012B, Marulli_2012A, delaTorre_2013, Beutler_2014, Alam_2017, Sanchez_2017, Satpathy_2017, pezzotta2017}. In this paper we investigate the spatial properties of the LSS of the Universe focusing on MG models based on the \citet{Hu_Sawicki_2007} $f(R)$ gravity. It is well known that the large-scale velocity field, as captured by e.g. the velocity power spectrum, is more sensitive to modifications of gravity as compared to the matter density distribution \citep{Jennings_RSD_MG_2012} and can therefore increase the signal associated with a deviation from standard GR. In the recent work by \citet{hagstotz2019}, some kinematic information encoded by the velocity power spectrum and by the velocity dispersion around massive clusters as extracted from a new suite of MG cosmological simulations -- the \dustp simulations \citep{Giocoli_lensing_2018A} -- was employed for the first time to disentangle the f(R)-massive neutrino degeneracy. In the present work we investigate the information gain coming from the large-scale velocity field through the redshift-space distribution of biased tracers, such as CDM haloes expected to host galaxies and galaxy clusters.
The current work follows from the analysis presented in \citet{Marulli_2012A}, who investigated the real-space and redshift-space clustering properties of CDM halo catalogues extracted from N-body simulations assuming coupled dark energy (cDE) models. Here we extend the latter analysis by exploiting the \dustp simulations, to explore the possible cosmic degeneracies introduced by a hot DM component consisting of massive neutrinos. Moreover, the statistical methodology is significantly updated, to closely match the one currently used to analyse real datasets. As in \citet{Marulli_2012A}, we focus on the redshift-space anisotropic two-point correlation function (2PCF) of CDM haloes, in real and redshift space. However, instead of analysing the two-dimensional (2D) 2PCF, we will perform a joint analysis of the monopole and quadrupole moments. Moreover, differently from \citet{Marulli_2012A}, we will perform a full Monte Carlo Markov Chain (MCMC) statistical analysis to derive posterior constraints on cosmological parameters, with a full covariance matrix estimated with bootstrap (instead of just diagonal Poisson errors), and extracting information on the linear bias of CDM haloes and on the linear growth factor in terms of $b\sigma_8$ and $f\sigma_8$ (instead of just measuring the linear distortion parameter, $\beta$).

The paper is structured as follows. In Section~\ref{sec:MGmodels} we summarise the theoretical framework of the \citet{Hu_Sawicki_2007} $f(R)$ model used in our work, and the cosmological effects of massive neutrinos. In Section~\ref{sec:simulations} we introduce the set of \dustp N-body simulations and the selected CDM halo samples, while the analyses of real-space and redshift-space clustering are presented in Sections \ref{sec:xiRS} and \ref{sec:xiZS}, respectively. The modelling of dynamic redshift-space distortions and the derivation of the parameter constraints are described in Section~\ref{sec:modelling}. Finally, in Section~\ref{sec:conclusions} we summarise the main results of this work and draw our conclusions.

\section{Modified gravity models and massive neutrinos}\label{sec:MGmodels}
Among the proposed extensions of GR, we consider the one based on the following modified Einstein-Hilbert action:
\begin{equation}
\label{fRaction}
S = \int {\rm d}^4x \, \sqrt{-g} \left( \frac{R+f(R)}{16 \pi G} + {\cal L}_m \right) \, ,
\end{equation}
where $R$ is the Ricci scalar, $G$ is the Newton's gravitational constant, $g$ is the determinant of the metric tensor $g_{\mu\nu}$, and ${\cal L}_m$ is the Lagrangian density of all matter fields\footnote{We use natural units $c=1$. The Greek indices, $\mu$ and $\nu$, run over $0$, $1$, $2$, $3$.}. A plausible $f(R)$ function able to satisfy the solar system constraints and, at the same time, to mimic the $\Lambda$CDM expansion history of the Universe, is given by:
\begin{equation}\label{fRHS}
f(R) = -m^2 \frac{c_1 \left(\frac{R}{m^2}\right)^n}{c_2 \left(\frac{R}{m^2}\right)^n + 1} \, ,
\end{equation}
where the mass scale $m$ is defined as $m^2 \equiv H_0^2 \Omega_{\rm M}$, and $c_{1}$, $c_{2}$ and $n$ are non-negative free parameters of the model \citep{Hu_Sawicki_2007}. For this $f(R)$ model, the background expansion history is consistent with the $\Lambda$CDM case by choosing $c_{1}/c_{2} = 6\Omega _{\Lambda}/\Omega _{\rm  M}$, where \Ol and \Om are the dimensionless density parameters for vacuum and matter, respectively. The scalar field $f_{R}\equiv df(R)/dR$ adds an additional degree of freedom to the model, whose dynamic in the limit of $|f_R| \ll 1$ and $|f/R|\ll 1$ can be expressed in terms of perturbations of the scalar curvature, $\delta R$, and matter density, $\delta\rho$:
\begin{equation}\label{eq:modPoisson}
\nabla^2 f_R = \frac{1}{3}\left(\delta R - 8 \pi G \delta \rho \right)\,.
\end{equation}
\begin{table*}
\caption{Summary of parameters used in the \dustp simulations considered in this work: $f_{R0}$ represents the modified gravity parameter, $m_\nu$ is the neutrino mass in Electronvolt, $\Omega_{\rm CDM}$ and $\Omega_{\nu}$ are the $\mathrm{CDM}$ and neutrino density parameters, $m^p_{\rm CDM}$ and $m^p_{\nu}$ are the CDM and neutrino particle masses (in M$_{\odot }/h$), respectively. The value in the last column displays the $\sigma_8$ parameter at $z=0$, which corresponds to the linear density fluctuations smoothed on a scale of 8\Mpch, computed from linear theory.}
\begin{tabular}{lcccccccc}
Simulation name & Gravity model  &  
$f_{R0} $ &
$m_{\nu }$ [eV] &
$\Omega _{\rm CDM}$ &
$\Omega _{\nu }$ &
$m^{p}_{\rm CDM}$ [M$_{\odot }/h$] &
$m^{p}_{\nu }$ [M$_{\odot }/h$] & $\sigma_8$ \\
\hline
$\Lambda $CDM & GR & -- & 0 & 0.31345 & 0 & $8.1\times 10^{10}$  & 0 & $0.842$ \\ 
$fR4$ & $f(R)$  & $-1\times 10^{-4}$ & 0 & 0.31345 & 0 & $8.1\times 10^{10}$  & 0 & $0.963$ \\ 
$fR5$ & $f(R)$  & $-1\times 10^{-5}$ & 0 & 0.31345 &0  & $8.1\times 10^{10}$  & 0 & $0.898$ \\ 
$fR6$ & $f(R)$  & $-1\times 10^{-6}$ & 0 & 0.31345 & 0 & $8.1\times 10^{10}$  & 0 & $0.856$ \\ 
$fR4\_0.3eV$ & $f(R)$  & $-1\times 10^{-4}$ & 0.3 & 0.30630 & 0.00715 & $7.92\times 10^{10}$ & $1.85\times 10^{9}$ & $0.887$ \\ 
$fR5\_0.15eV$ & $f(R)$  & $-1\times 10^{-5}$ & 0.15 & 0.30987 & 0.00358 & $8.01\times 10^{10}$ & $9.25\times 10^{8}$ & $0.859$ \\ 
$fR5\_0.1eV$ & $f(R)$  & $-1\times 10^{-5}$ & 0.1 & 0.31107 & 0.00238 & $8.04\times 10^{10}$ & $6.16\times 10^{8}$ & $0.872$ \\ 
$fR6\_0.06eV$ & $f(R)$  & $-1\times 10^{-6}$ & 0.06 & 0.31202 & 0.00143 & $8.07\times 10^{10}$ & $3.7\times 10^{8}$  & $0.842$ \\
$fR6\_0.1eV$ & $f(R)$  & $-1\times 10^{-6}$ & 0.1 & 0.31107 & 0.00238 & $8.04\times 10^{10}$ & $6.16\times 10^{8}$ & $0.831$ \\ 
\hline
\end{tabular}
\label{tab:models}
\end{table*}
\begin{figure*}
  \includegraphics[width=\hsize]{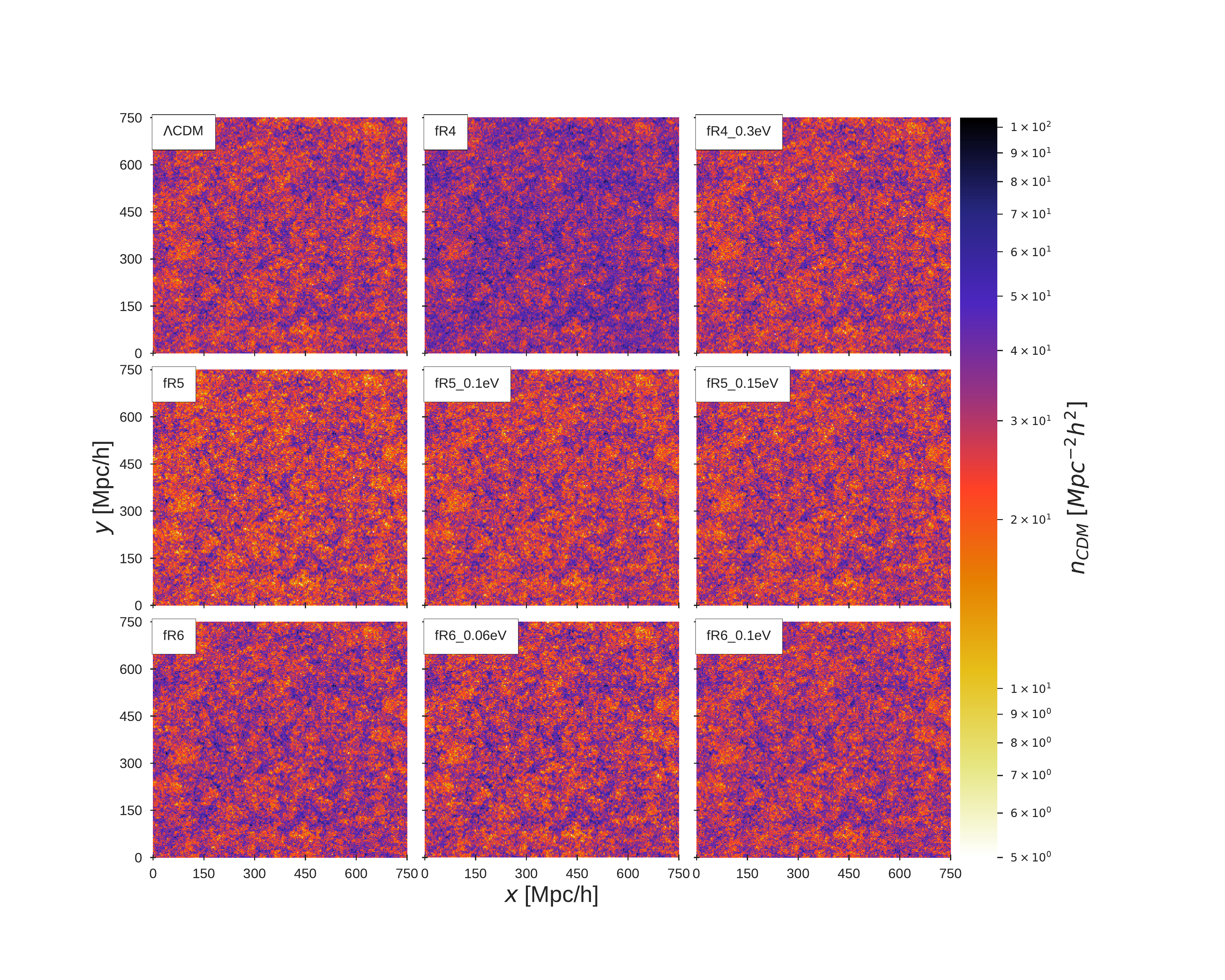}
  \caption{Maps of the projected number density of CDM haloes in the mass range $\left[4\times10^{12},\;7\times10^{14}\right]{\mbox M}_\odot/h$ extracted from the \dustp simulations at $z=0$. The boxes have been divided into $300\times300$ pixels and the colorbar indicates the normalised number of CDM haloes $n_{\rm CDM}$ per unit area ($2.5\times2.5~\mbox{Mpc}^2h^{-2}$).
    \label{fig:DensityBoxes}}
\end{figure*}
Comparing to the $\Lambda$CDM model expansion history, and under the condition $c_2 (R/m^2)^n  \gg 1$, the scalar field can be approximated by: 
\begin{equation}\label{eq:fR-R_relation}
f_R \approx -n \frac{c_1}{c_2^2}\left(\frac{m^2}{R}\right)^{n+1}\,.
\end{equation}
Thus, for $n=1$ the model is fully specified by only one free parameter $c_2$, which in turn can be expressed in terms of 
the dimensionless scalar at present epoch, $f_{R0}$, given by:
\begin{equation}
f_{R0}\equiv -\frac{1}{c_{2}}\frac{6\Omega _{\Lambda }}{\Omega_{M}}\left( \frac{m^2}{R_{0}}\right) ^{2}\,.
\end{equation}
Under these assumptions, the modified Einstein's field equations for $f(R)$ gravity lead to a dynamical gravitational potential, $\Phi = \Phi_N - \delta R/6$, that satisfies the following equation:
\begin{equation}\label{eq:fRPoisson}
\nabla^2 \Phi = -\frac{16\pi G}{3}\delta\rho - \frac16\delta R\,,
\end{equation}
being $\Phi_N$ the Newtonian potential. 

Massive neutrinos suppress the clustering below their thermal free-streaming scale and change the matter-radiation equality time \citep{Lesgourgues_neutrinos_2006}. They also affect the non-linear matter power spectrum \citep{Brandbyge_2008, Saito_2008, Saito_2009, Brandbyge_2009, Brandbyge_2010, Agarwal_2011, Wagner_2012}, the halo mass function \citep{Brandbyge_2010_halos, Marulli_neutrino_2011, Villaescusa_cosmicbackground_2013a}, the clustering properties of CDM haloes and redshift-space distortions \citep{Viel_neutrinoeffect_2010, Marulli_neutrino_2011, villaescusa2014, Castorina_neutrino_2014, Castorina_2015, Zennaro_neutrinos_2018}, and the scale-dependent bias \citep{Chiang_scaledepbias_2018arXiv}. As they are non-relativistic particles at late times, massive neutrinos contribute to the total energy density of the Universe $\Omega_M$, so that $\Omega_M=\Omega_{\rm CDM}+\Omega_b+\Omega_\nu$, where $\Omega_{\rm CDM}$ and $\Omega_b$ are the dimensionless density parameters for CDM and baryons, respectively, and the contribution related to the massive neutrino component, $\Omega_\nu$, can be expressed in terms of the total neutrino mass, $m_{\nu}\equiv \sum_im_{\nu_i}$, as follows:
\begin{equation}\label{eq:Omeganu}
\Omega_\nu=\frac{\Sigma_i m_{\nu_i}}{93.14~h^2\mbox{eV}}\, .
\end{equation}
Several astronomical observations provide upper limits on the total neutrino mass, that results around $0.1-0.3$ eV, under the assumption of standard GR \citep[see e.g.][]{seljack2006, Riemer_2013, Jia-Shu_2015, Lu2016, Cuesta_2016, Kumar_2016, Yeche_2017, Poulin_2018}.

However, one of the most important goals of current cosmology is to extract robust, model-independent constraints on neutrino masses. It is thus crucial to investigate whether the cosmological effects of massive neutrinos might be degenerate with MG models, which would severely affect the constraints. Cosmological probes able to distinguish between these two effects are required to achieve tight constraints on both MG and massive neutrinos \citep{He_neutrinos_2013, Motohashi_neutrinos_2013, Baldi_degeneracies_2014, Bellomo_Hidingneutrino_2017, Wright_COLA_2017, Peel_degeneracy_2018, Giocoli_lensing_2018A}. In the following sections, we will address this issue focusing in particular on the halo clustering, modelled through DM N-body simulations that include simultaneously both effects.
\begin{figure*}
  \includegraphics[width=0.995\textwidth]{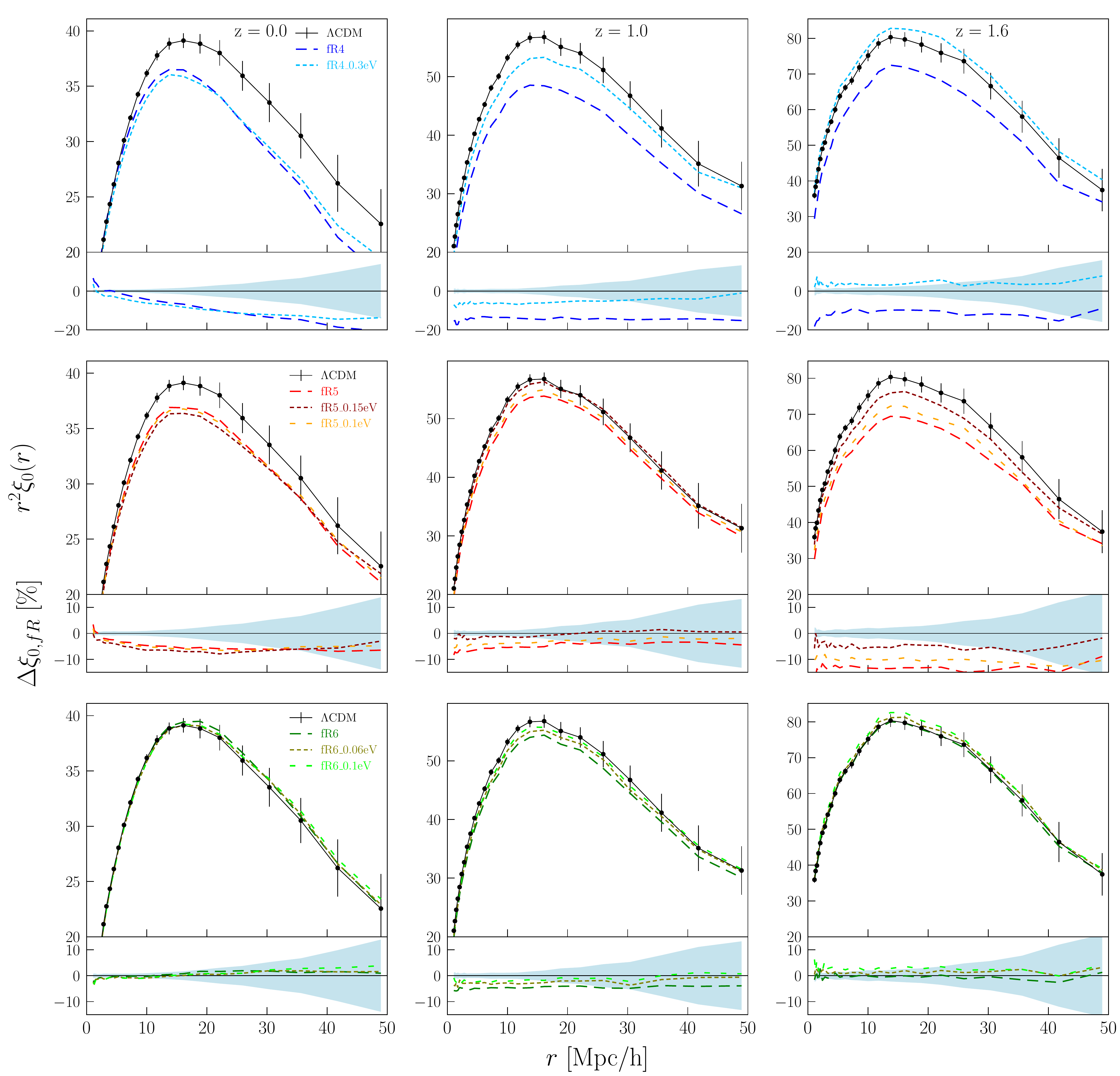}
  \caption{The real-space 2PCF $\xi_0$ of CDM haloes for all the models of the \dustp project at three different redshifts: $z=0$ (left column), $z=1$ (central column), $z=1.6$ (right column). From top to bottom, the panels show the $fR4$, $fR5$ and $fR6$ models, respectively, compared with the results of the $\Lambda$CDM model. The error bars, shown only for the $\Lambda$CDM model for clarity reasons, are the diagonal values of the bootstrap covariance matrices used for the statistical analysis. Percentage differences between $f(R)$, $f(R)+m_\nu$ and $\Lambda$CDM predictions are in the subpanels, while the shaded regions represent the deviation at 1$\sigma$ confidence level.
  \label{fig:multipolesRS}}
\end{figure*}

\section{N-body simulations and halo samples} \label{sec:simulations}
We use a subset of the \dustp (Dark Universe Simulations to Test GRAvity In the presence of Neutrinos) cosmological N-body simulations, which is part of a numerical project aimed at investigating possible cosmic degeneracies, such as the ones between $f(R)$ models and massive neutrinos, that is the subject of the present work. In a series of recent papers, these simulations have been exploited to investigate several features related to weak-lensing statistics \citep{Giocoli_lensing_2018A, Peel_degeneracy_2018}, to the abundance of massive haloes \citep{Hagstotz_2018} and to explore cosmic degeneracies using machine learning techniques \citep{Peel_2018_MG_MachiLearn, Merten_2018_dissection}. 

The \dustp runs have been performed using the {\small MG-Gadget} code \citep{Puchwein_2013}, which is a modified version of {\small GADGET} \citep{Springel_GADGET2_2005} implementing the \citet{Hu_Sawicki_2007} $f(R)$ gravity model, with a mixture of cold and hot DM components, the latter made up of massive neutrinos. The simulations were carried out in a box of $(750$ Mpc$/h)^3$ volume, with periodic boundary conditions, and $768^{3}$ DM particles. The cosmological parameters assumed for all the considered models at $z=0$ are consistent with Planck $2015$ constraints \citep{Planck_2015_XIII}: $\Omega_{\rm M}\equiv\Omega_{\rm CDM}+\Omega_{\rm b}+\Omega_{\nu} = 0.31345$, $\Omega_{\rm b}=0.0481$, $\Omega_{\Lambda}= 0.68655$, $H_{0}= 67.31$ km s$^{-1}$ Mpc$^{-1}$, ${\cal{A}} _{\rm  s}= 2.199\times 10^{-9}$, $n_{s}=0.9658$ and  $\sigma_8=0.847$. 

We identify haloes in the particle distribution using the Spherical Overdensity (SO) algorithm termed \texttt{Denhf} \citep{tormen98a,tormen04,giocoli08a,despali16}. We chose this method over the Friends-of-Friends (FoF) group finding algorithm by \citet{Davis_FOF_1985}, because it appears to be slightly closer to physical models of halo formation, and because of its resemblance to the definition of the mass in observational data sets. Specifically, for each particle we compute the local DM density by calculating the distance $d_{i,10}$ to the tenth nearest neighbour. In this way, we assign to each particle a local density $\rho_i \propto d_{i,10}^{-3}$. Next, we sort the particles by density and define the position of the densest particle as the centre of the first halo. Around this centre, the algorithm grows a sphere with a certain average density, that in this work has been chosen to be $200$ times the critical density of the Universe. At this point we assign all particles within the sphere to the newly identified halo, removing them from the global list of particles. Subsequently, the densest particle of the remaining distribution is chosen and the process is repeated several times, until none of the remaining particles has a local density large enough to be the centre of a 10 particle halo. %Particles not assigned to any haloes are called  `field' or `dust' particles. 
In numerical simulations containing massive neutrinos, we assume that they contribute only to the expanding cosmological background metric \citep{Castorina_neutrino_2014} and thus, when identifying the haloes, we link together only DM particles.

Tab.~\ref{tab:models} presents an overview of the main parameters of each simulation, such as the $f_{R0}$ values, the total neutrino mass, the total CDM density constrast, $\Omega_{\rm CDM}$, and the mass of the DM particles. In all cases, the scalar at present epoch, $|f_{R0}|$, is in the range $10^{-4}-10^{-6}$, as suggested by \citet{Hu_Sawicki_2007}, to be consistent with distance-based measurements of the expansion history. The total neutrino masses considered in this work are $m_\nu=0,~0.06,~0.1,~0.15,~0.3$ eV. As shown in Fig.~\ref{fig:DensityBoxes} at $z=0$, the density distributions of CDM haloes predicted by the MG models considered show notable differences, as it can be appreciated, for instance, comparing $f(R)$ and $f(R)+m_\nu$ models.

For the clustering analysis presented in the following Sections, we make use of halo samples from each of the nine models presented in Tab.~\ref{tab:models}, restricting our analysis in the mass range $M_{\rm min}<M<M_{\rm max}$, where $M_{\rm min}=4\times10^{12} \mbox{M}_\odot/h$ and $M_{\rm max}=7\times10^{14},~4\times10^{14},~3\times10^{14},~2\times10^{14},~10^{14} \mbox{M}_\odot/h$ at $z=0,~0.5,~1,~1.4,~1.6$, respectively. 

\section{Clustering in real space}\label{sec:xiRS}
In this Section, we describe the methodology used to quantify the halo clustering in real space, focusing on the first multipole moment of the 2PCF, that is the monopole. All the numerical computations in the current Section and in the following ones have been performed with the {\small CosmoBolognaLib}, a large set of {\em free software} libraries that provide all the required tools for the data analysis presented in this work, including the measurements of all statistical quantities and the Bayesian inference analysis\footnote{Specifically, we used {\small CosmoBolognaLib V5.0}. The CosmoBolognaLib are entirely implemented in C++. They also provide the possibility to be converted in Python modules through wrappers. Both the software and its documentation are freely available at the public GitHub repository: \href{https://github.com/federicomarulli/CosmoBolognaLib}{https://github.com/federicomarulli/CosmoBolognaLib}.} \citep{Marulli_CosmoBologna}.

\subsection{The two-point correlation function}\label{subsec:xiRS}
We measure the 2D 2PCF, $\xi(r,\mu)$, with the  \citet{LandySzalay_1993} estimator given by:
\begin{equation}
\hat{\xi}(r,\mu)=\frac{DD(r, \mu)-2DR(r, \mu) +RR(r, \mu)}{RR(r, \mu)} \, ,
\end{equation}
where $\mu$ is the cosine of the angle between the line of sight and the comoving halo pair separation, $r$, and $DD(r,\mu), RR(r,\mu)$, and $DR(r,\mu)$ represent the normalised number of data-data, random-random and data-random pairs, respectively, in ranges of $r$ and $\mu$. We consider intermediate scales from $1$\Mpch~to $50$\Mpch, in $25$ logarithmic bins. The random samples used are ten times larger than the halo ones. The 2PCF uncertainties are estimated with the bootstrap method, by dividing the original data sets into $27$ sub-samples, which are then re-sampled in $100$ data sets with replacement, measuring $\xi(r,\mu)$ in each one of them \citep{Efron_1979, Barrow_1984, Ling_1986}. 

It is convenient to expand the 2D 2PCF in terms of Legendre polynomials, $L_l(\mu)$, as follows:
\begin{equation}\label{eq:ximultiexp}
\xi(s,\mu)\equiv \xi_0(s)L_0(\mu)+\xi_2(s)L_2(\mu)+\xi_4(s)L_4(\mu)\, ,
\end{equation}
where each coefficient corresponds to the $l^{th}$ multipole moment:
\begin{equation}
\xi_l(r) = \frac{2l +1}{2}\int_{-1}^{+1} d\mu\, \xi(r,\mu) L_l(\mu)\, .
\end{equation} 
The clustering multipoles are computed with the \emph{integrated estimator} \citep[e.g.][]{Kazin_2012}, which consists in calculating $\xi(r,\mu)$ in 2D bins and then integrating it as follows:
\begin{equation}\label{eq:integratedxil}
\hat{\xi}_l(r) = \frac{2l +1}{2}\int_{-1}^{+1} d\mu L_l(\mu) \frac{DD(r, \mu)-2DR(r, \mu) +RR(r, \mu)}{RR(r, \mu)}\,.
\end{equation}

In real space the full clustering signal is contained in the monopole moment $\xi_0(r)$. Fig.~\ref{fig:multipolesRS} shows $\xi_0(r)$ of CDM haloes for all models considered in the \dustp project, at three different redshifts $z=0, 1, 1.6$. Subpanels show the percentage difference between MG models [$f(R)$ with and without massive neutrinos], and the $\Lambda$CDM model, computed as $\Delta\xi_{\rm fR}=100(\xi_{\rm fR}-\xi_{\rm \Lambda CDM})/\xi_{\rm \Lambda CDM}$.

The clustering properties of $fR4$ and $fR4\_0.3eV$ models at $z=0$ are the ones that deviate the most from $\Lambda$CDM, with a significant clustering suppression at scales larger than $10$ \Mpch. This is expected, as the $f_{R0}$ value of these models is the most extreme one considered, marginally compatible with the constraints from solar system observations \citep{Hu_Sawicki_2007}. At higher redshifts the $fR4\_0.3eV$ monopole gets closer to the $\Lambda$CDM one, due to the effect of massive neutrinos. A similar, though less significant, effect is found also for $fR5$, $fR5\_0.15eV$ and $fR5\_0.1eV$ models. The clustering suppression is further reduced in the $fR6$, $fR6\_0.06eV$ and $fR6\_0.1eV$ models, so that they appear highly degenerate with $\Lambda$CDM at all the scales and redshifts considered, with deviations smaller than $2\%$.

\begin{figure*}
    \includegraphics[width=0.9\hsize]{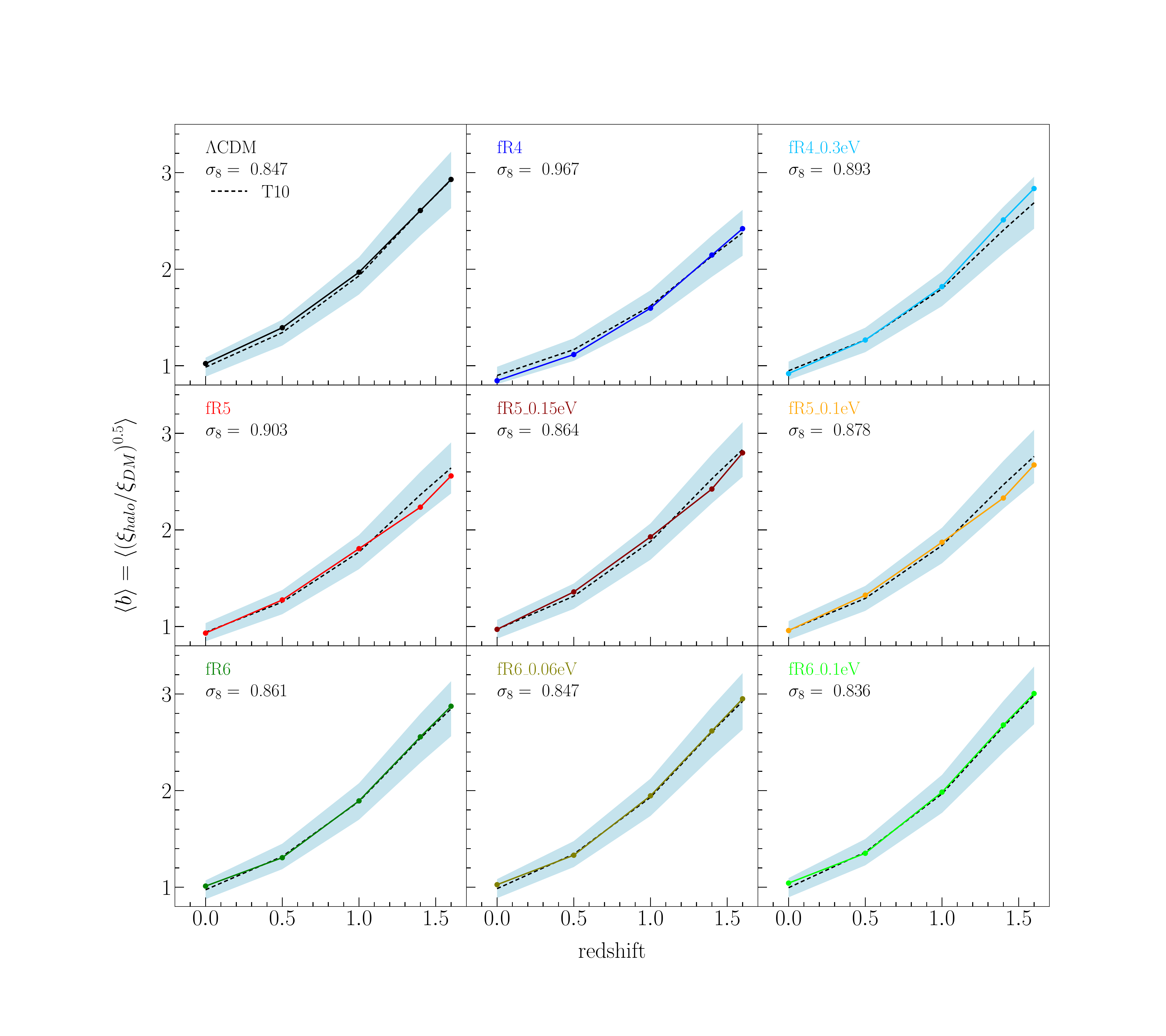}
	\caption{The coloured solid lines represent the {\em apparent} effective halo bias, $\langle b\rangle$, as a function of redshift, averaged in the range 10\Mpch$<r<$50\Mpch. Black lines show the theoretical $\Lambda$CDM effective bias predicted by \citet{Tinker_bias_2010} (dashed), normalised to the $\sigma_{8}$ values of each \dustp simulation, while the cyan shaded areas show a 10\% error.
	\label{fig:bias_redshift}}
\end{figure*}

\subsection{The halo biasing function}\label{subsec:bias}
\begin{figure*}
	\includegraphics[width=\hsize]{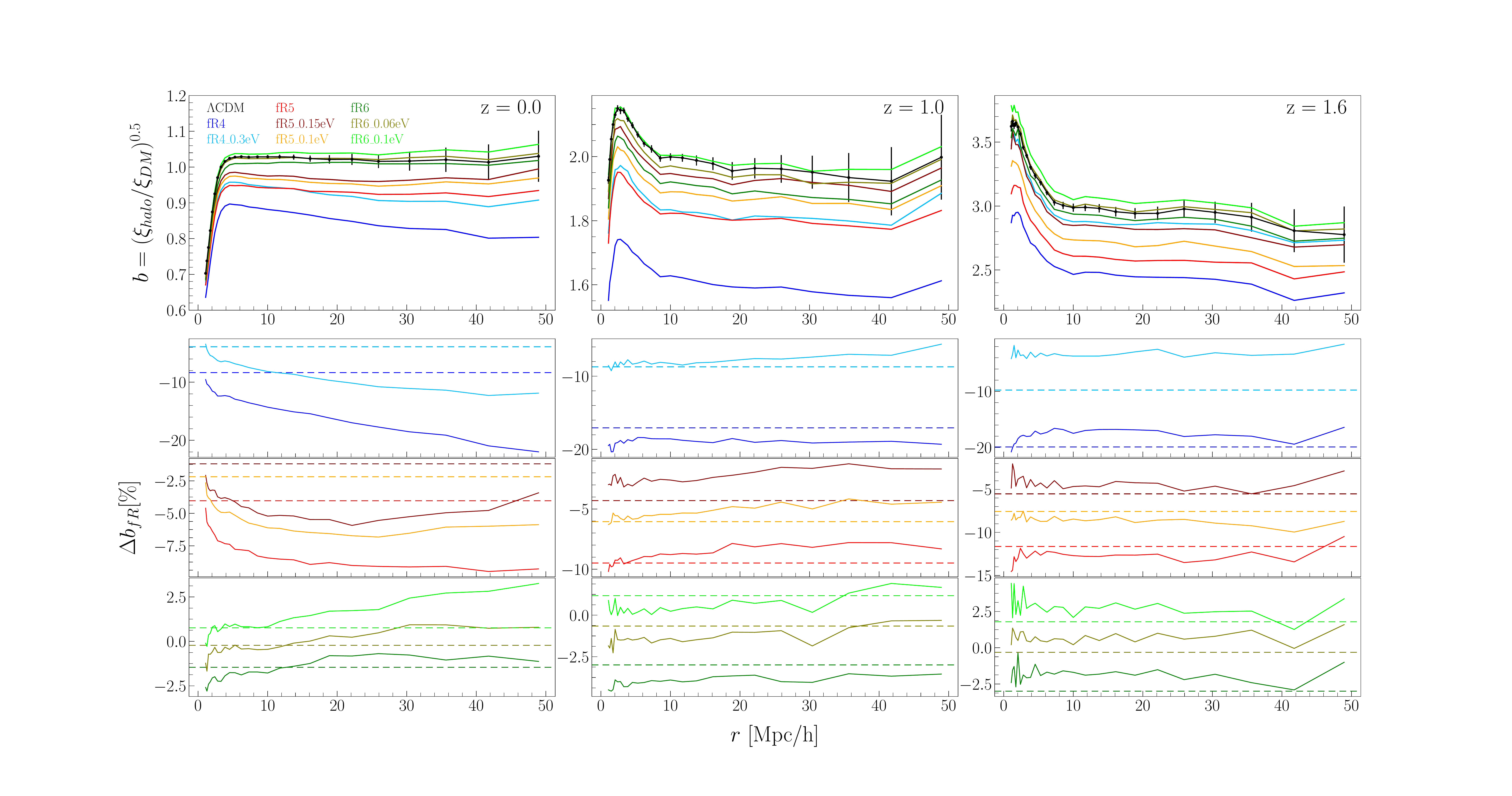}
	\caption{The {\em apparent} effective halo bias as a function of scale, for three different redshifts ($z=0,~1$ and $1.6$, columns from left to right), for all models, as indicated by the labels. Dotted and dashed horizontal lines show theoretical predictions, as in Fig. \ref{fig:bias_redshift}.
	\label{fig:bias_scale}}
\end{figure*}

To characterise the relation between the halo clustering and the underlying mass distribution, we estimate the effective halo bias. In the linear regime, the bias is approximately independent of the scale, depending only on halo masses and redshifts. In our mass-selected samples, this quantity can be computed as follows, averaging in a given scale range:
\begin{equation}\label{eq:biasefflss} 
\left\langle b(z)\right\rangle=\left\langle\sqrt{\frac{\xi_{\rm halo, fR}}{\xi_{{\rm DM},\Lambda {\rm CDM}}}}~\right\rangle, 
\end{equation}
where $\xi_{\rm halo, fR}$ and $\xi_{{\rm DM},\Lambda {\rm CDM}}$ are the CDM halo 2PCF of the \dustp models and the CDM 2PCF estimated in $\mathrm{\Lambda CDM}$, respectively. Eq. \eqref{eq:biasefflss} is obtained by Fourier transforming the non-linear matter power spectrum computed with {\small CAMB}, including {\small HALOFIT} \citep{Lewis_2000, Smith_halofit_2003}. The latter simulates the {\em apparent} bias that would be assessed in a $f(R)$ Universe if a $\Lambda$CDM model was wrongly assumed to predict the DM clustering \citep[see][for more details]{Marulli_2012B}. The {\em apparent} effective bias is then estimated by averaging the bias $b(M,z)$ over a set of CDM haloes with given mass $M_i$:
\begin{equation}\label{eq:biaseff}
b(z) = \frac{1}{N_{\rm halo}}\sum_{i=1}^{N_{\rm halo}} b(M_i, z)\, .
\end{equation} 
In order to compare measurements in $f(R)$ and $f(R)+m_\nu$ scenarios with the $\Lambda$CDM ones, we consider the theoretical effective bias proposed by \citet{Tinker_bias_2010}, computed with the so-called {\em CDM prescription} \citep{villaescusa2014}, that is, using the linear CDM+baryons power spectrum\footnote{Both $P^{\rm CDM+b}_{lin}(k)$ and $P^{m}_{lin}(k)$ can be directly obtained with {\small CAMB}, since $P^{\rm CDM}_{lin}(k)= T_{\rm CDM}^2/T_m^2 P^m_{lin}(k)$, where $T_{\rm CDM}(k)$ and $T_b(k)$ are the corresponding transfer functions.}, and replacing $\rho_m$ with $\rho_{\rm CDM}$ \citep{Castorina_neutrino_2014}. The CDM prescription has, however, a minor impact on the results presented in this work. Comparing to the results obtained with the total matter power spectrum, we found deviations on the estimated halo bias smaller than $1\%$ for $fR5+m_\nu$ and $fR6+m_\nu$ models, and of about $3\%$ for $fR4+m_\nu$ model. Fig.~\ref{fig:bias_redshift} shows the mean {\em apparent} effective bias as a function of redshift, averaged over the range 10\Mpch$<r<$50\Mpch, whereas Fig.~\ref{fig:bias_scale} shows how it changes as a function of scale. The error bars are computed by propagating the 2PCF uncertainties obtained with the bootstrap method (see Section \ref{subsec:xiRS}). Dashed lines represent the theoretical expectations by \citet{Tinker_bias_2010}, while the shaded region shows a $10\%$ difference with respect to the central value. The effective bias increases as a function of redshift, as expected \citep{Matarrese_1997, Chung-Pei_1999}. The predicted effective bias of all the models considered appears quite indistinguishable from the $\Lambda$CDM case, when it is normalised to the $\sigma_8$ values of the \dustp cosmologies, that is assuming $\xi_{DM,f(R)}=\xi_{DM,\Lambda {\rm CDM}(\sigma_{8})}$, where $\xi_{DM,\Lambda {\rm CDM}(\sigma_{8})}$ is computed by setting the amplitude of the primordial curvature perturbations to the values required to have the $\sigma_8$ values of the $f(R)$ models \citep[see e.g.][]{Marulli_neutrino_2011, Marulli_2012A}. The largest deviation occurs at $z=1.6$ for the $fR4\_0.3eV$ model, though it is in any case not statistically significant (between $5\%$ and $7\%$, averaging over different distances larger than $10$\Mpch). As a counterpart, the most degenerate model is $fR6$, both with and without massive neutrinos, which is in agreement with $\Lambda$CDM better than $2\%$ at all scales. 

%============================================
\section {Clustering in redshift-space}\label{sec:xiZS}
Spectroscopic surveys observe a combination of density and velocity fields in redshift space. Specifically, the observed redshift, $z_{\rm obs}$, of extragalactic sources is a combination of the cosmological redshift, $z_c$, due to the Hubble flow, and an additional term caused by the peculiar velocities along the line of sight:
\begin{equation}\label{eq:zobs}
z_{\rm obs}=z_c+(1+z_c)\frac{\vec{v}\cdot\hat{x}}{c}\hat{x}\, , 
\end{equation}
where $\hat{x}$ is a unit vector along the line of sight, so that the contribution of peculiar motions is given by $\vec{v}_\parallel=\vec{v}\cdot\hat{x}$.
As a consequence, redshift-space catalogues appear distorted with respect to the real-space ones. Since in N-body simulations both positions and peculiar velocities are known, the distorted mass distribution in redshift space can be derived directly. Specifically, we first convert the comoving coordinates of each CDM halo, \{x, y, z\}, into polar real-space coordinates \{R.A., Dec, $z_c$\}, relative to a given virtual observer placed at random, where R.A. and Dec are the Right Ascension and Declination, respectively. Then, we estimate the observed redshifts using Eq. \eqref{eq:zobs}. Finally, we convert back \{R.A., Dec, $z_{\rm obs}$\} into {\em distorted} comoving coordinates \{x', y', z'\}, mimicking the redshift space.

Redshift-space distortions turned out to be one of the most powerful cosmological probes to test the gravity theory on the largest scales \citep{Kaiser_1987, Guzzo_2008Natur,  Simpson_RSD_MG_PhysRevD_2010, Jennings_RSD_MG_2012, Raccanelli_RSD_2012, He_RSD_2018}. In redshift space, the spatial statistics of cosmic tracers, such as the 2PCF and power spectrum, are anisotropic due to the dynamic distortions along the line of sight \citep{Hamilton_review_1998, Scoccimarro_review_2004}:  at large scales the matter density distribution appears squashed along the line of sight, while an opposite stretching distortion is present at small scales, the so-called {\em fingers of God} (FoG) effect \citep{Jackson_FoG_1972}.
\begin{figure}
  \includegraphics[width=\hsize]{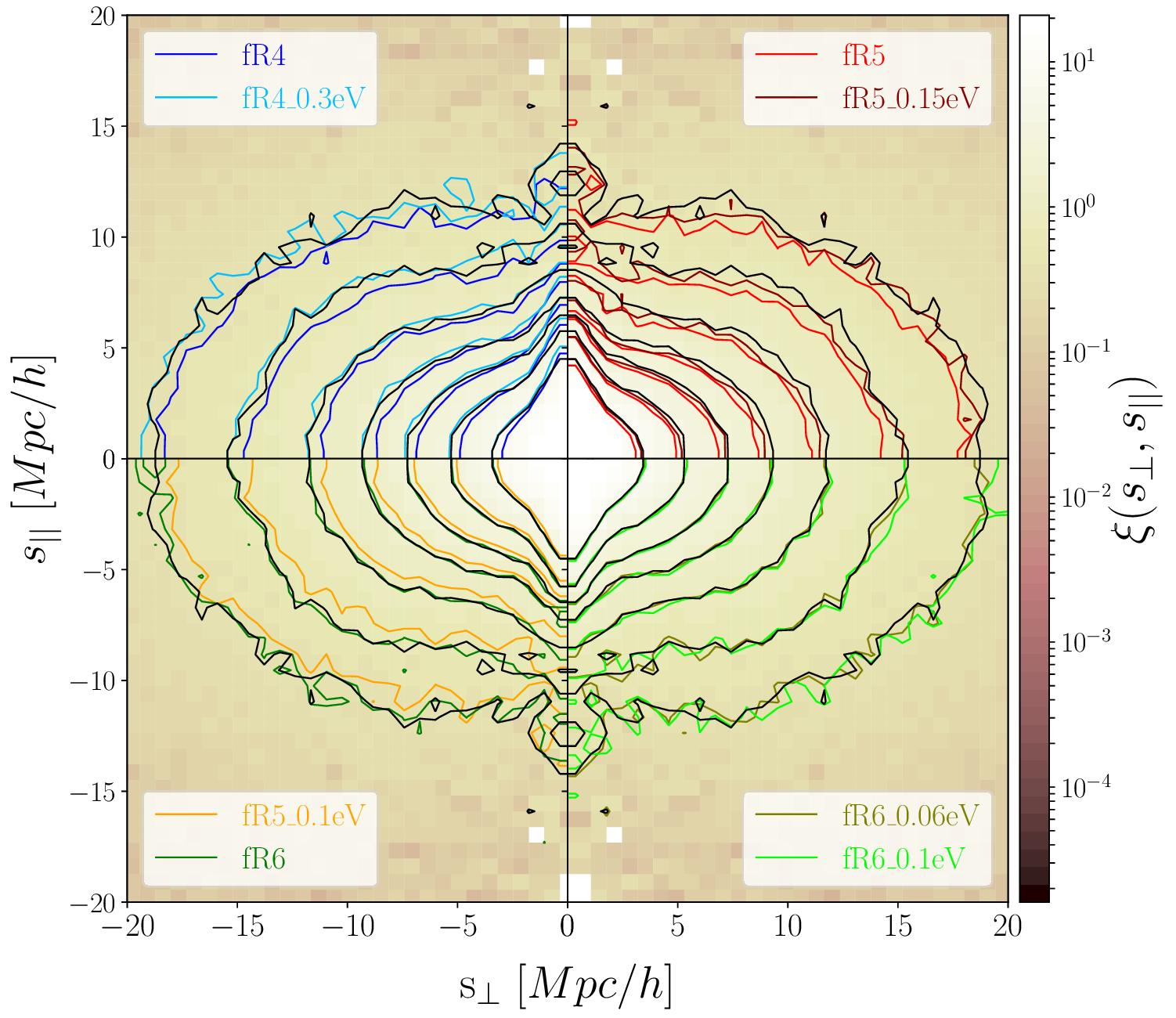}
  \caption{Contour lines of the 2D 2PCF of the \dustp simulations at $z=1.6$. Each quadrant refers to a different set of models, as labelled in the plot. The iso-curves plotted are \xiis$=\{0.3,~0.5,~1.0,~1.4,~2.2,~3.6,~7.2,~21.6\}$.
  \label{fig:FoG}}
\end{figure}
\begin{figure*}
  \includegraphics[width=\hsize]{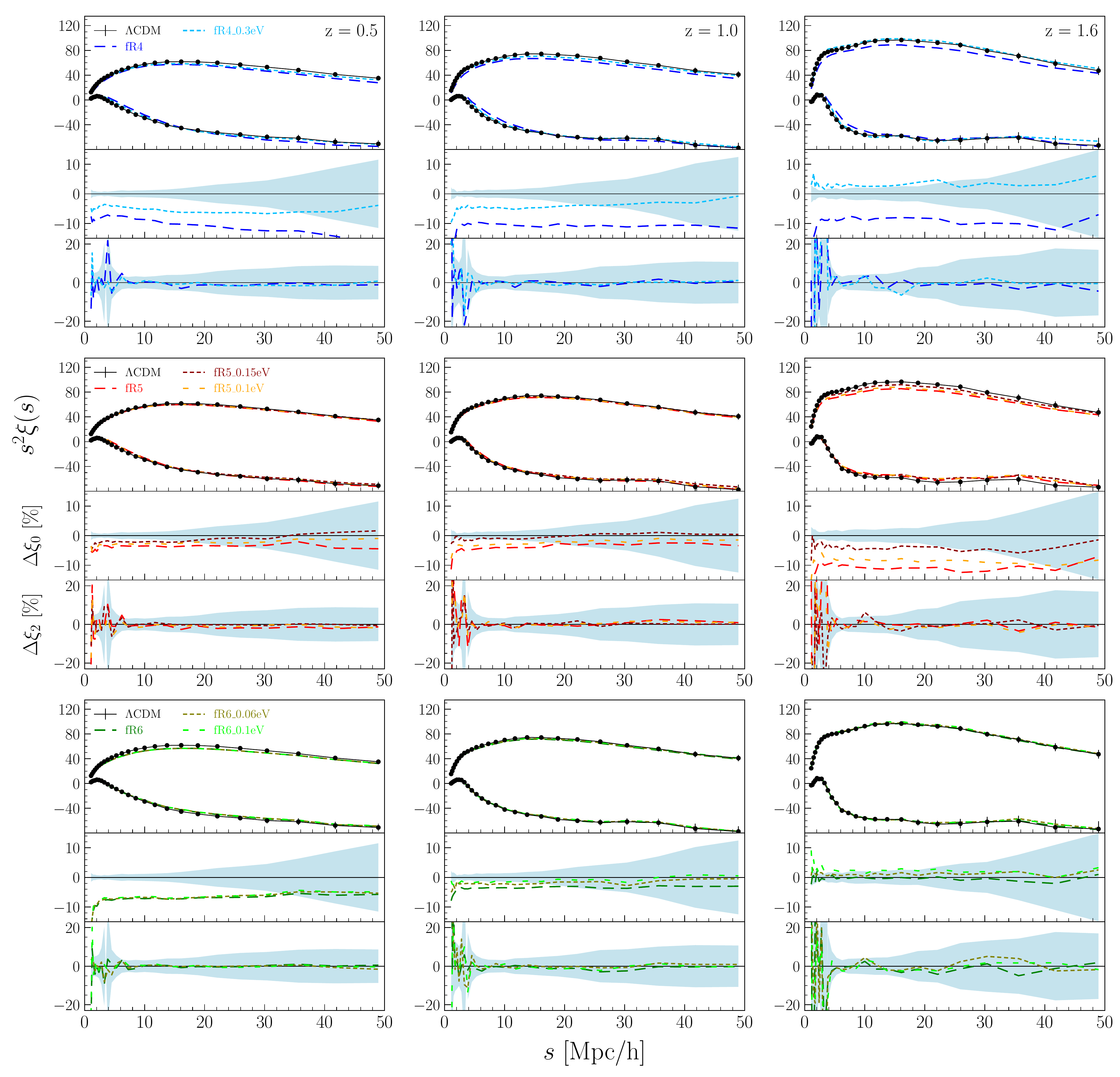}
  \caption{The redshift-space monopole (upper curves) and quadrupole (lower curves) moments of the 2PCF of the \dustp simulations at three different redshifts: $z=0.5$ (left column), $z=1$ (central column), and $z=1.6$ (right column). Black lines show the $\Lambda$CDM prediction compared to the results of different models (coloured lines, as labelled). The percentage differences between $f(R)$, $f(R)+m_\nu$ and $\Lambda$CDM predictions are in the subpanels. The cyan shaded regions represent the deviations at $\pm1\sigma$ confidence level.
  \label{fig:multipolesZS}}
\end{figure*}

The effect of redshift-space distortions on the 2PCF is shown conveniently by decomposing the pair comoving distances into their parallel and perpendicular components to the line of sight, that is $\vec{s}=(s_\parallel, s_\perp)$. Hereafter, we will use $s$ to indicate redshift-space coordinates. The anisotropic redshift-space 2PCFs, $\xi(s_\perp,s_\parallel)$, of all our \dustp halo catalogues at $z=1.6$ are shown in Fig. \ref{fig:FoG}.
Similarly to the real-space case, the 2PCF predicted by the $f(R)$ model with $|f_{R0}|=10^{-6}$, both with and without massive neutrinos, is quite similar to the $\Lambda$CDM one. On the other hand, the $|f_{R0}|=10^{-4}$ model with massless neutrinos predicts a lower signal on all scales. 

As described in Section \ref{subsec:xiRS}, the 2D 2PCF can be conveniently expressed in the Legendre multipole base. We focus here only on the first two even multipoles of the 2PCF, that is the monopole, $\xi_0$, and the quadrupole, $\xi_2$. The signal in the other even multipoles of the redshift-space 2PCF of CDM halo in the considered simulations is negligible, while odd multipoles vanish by symmetry. 
Fig. \ref{fig:multipolesZS} shows the monopole and quadrupole of the redshift-space 2PCF of all the halo catalogues considered in this work. As expected, the  $\{|f_{R0}|=10^{-4},m_\nu=0.3eV\}$ model is the one that differs the most from $\mathrm{\Lambda CDM}$. This is particularly evident in the monopole. On the other hand, the quadrupole appears less sensitive to the effect of the alternative cosmologies considered, both with and without massive neutrinos. The percentage differences with respect to the $\Lambda$CDM case are shown in the lower panels. For the quadrupole they are always smaller than $5\%$, whereas for the monopole they can reach up to $10\%$.

\begin{figure*}
	\includegraphics[width=\hsize]{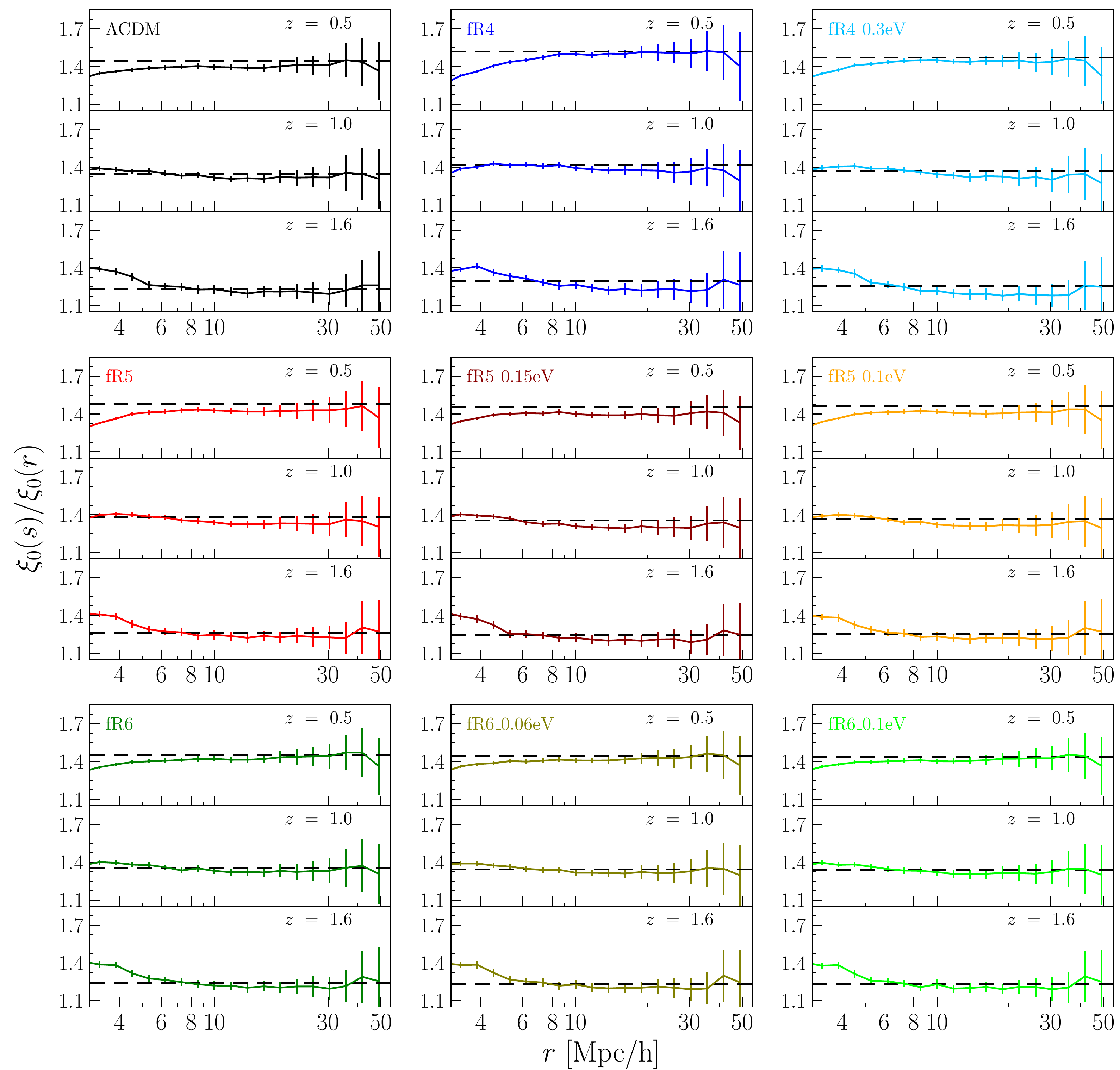}
	\caption{Ratio between the redshift-space and real-space 2PCF monopoles at redshifts $z=0.5$ (left column), $z=1$ (central column), $z=1.6$ (right column). From top to bottom, the panels show the results for the $fR4$, $fR5$ and $fR6$ models, respectively. Horizontal dashed lines show the theoretical predictions by \citet{Tinker_bias_2010}, normalised to the $\sigma_{8}$ values of each \dustp simulation.
	\label{fig:xi_ratio}}
\end{figure*}

\begin{figure}
  \includegraphics[width=\hsize, height=0.2\textwidth]{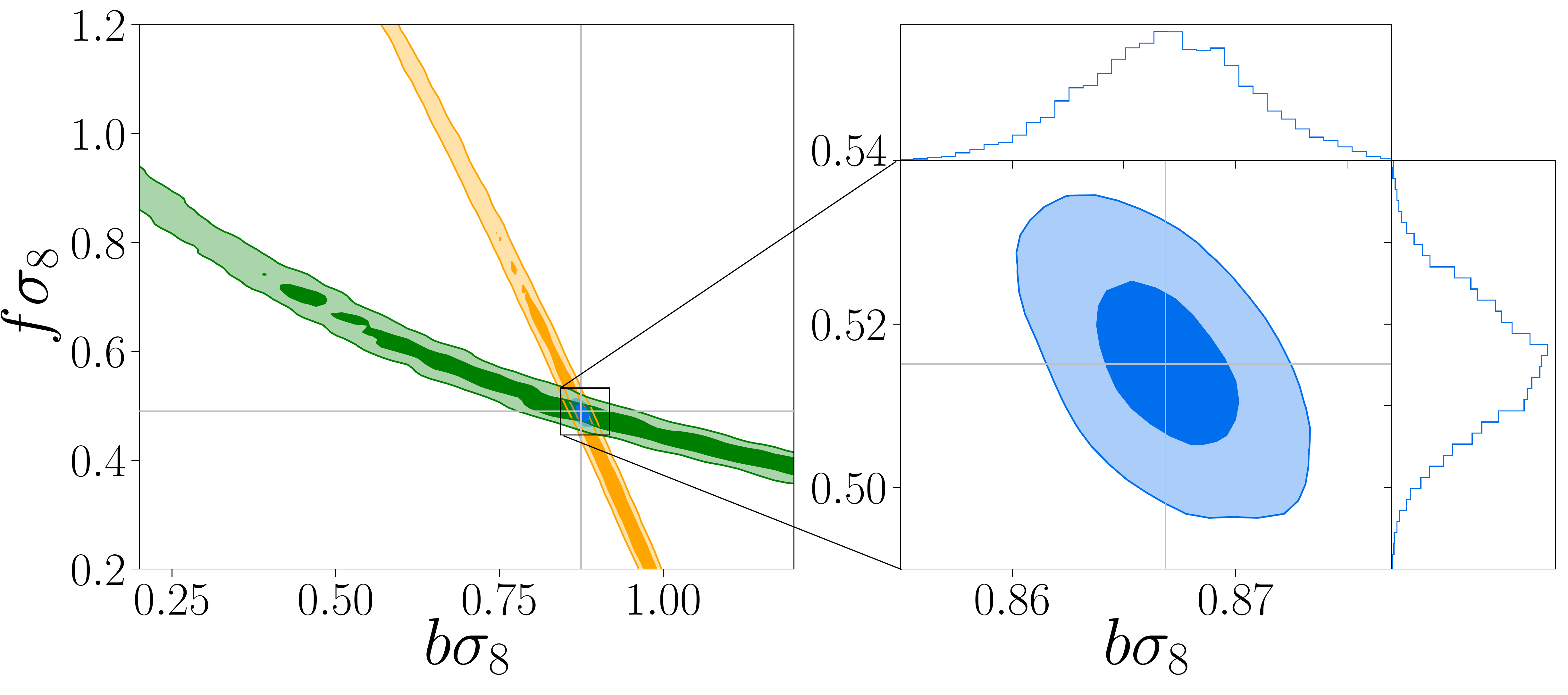}
  \caption{Contours at $1-2\sigma$ confidence level of the $f\sigma_8-b\sigma_8$ posterior distributions, obtained from the MCMC analysis in redshift-space for 2PCF multipoles of CDM haloes corresponding to the $\Lambda$CDM model at $z=0.5$: monopole (orange), quadrupole (green) and monopole plus quadrupole (blue). The joint modelling of monopole and quadrupole breaks the degeneracy in the $\{f\sigma_8,b\sigma_8\}$ space.
  \label{fig:breakingdegeneracy}}
\end{figure}

\section{Modelling the dynamic redshift-space distortions}\label{sec:modelling}
To quantify the effects of $f(R)$ gravity and massive neutrinos on redshift-space clustering distortions, we perform a statistical analysis aimed at extracting constraints on the growth rate of matter perturbation from the monopole and quadrupole of the redshift-space 2PCF of CDM haloes.
Following \citet{Marulli_2012A}, we start analysing the ratio between the redshift-space and real-space monopoles, which depends directly on the linear distortion parameter, $\beta$:
\begin{equation}\label{eq:xiratiomono}
\frac{\xi_0(s)}{\xi_0(r)} = 1 + \frac23\beta + \frac25\beta\,.
\end{equation}
The linear distortion parameter is defined as follows:
\begin{equation}
\beta\equiv\frac{f(\Omega_m)}{b}\simeq\frac{\Omega_m(z)^\gamma}{b}\,,
\end{equation}
where $f(\Omega_m)\equiv d\ln D/d\ln a$ is the linear growth rate, $D$ is the linear density growth factor, $b$ is the linear CDM halo bias and $\gamma$ is the gravitational growth index, which depends on the gravity theory. In GR, it can be demonstrated that $\gamma\sim0.545$ \citep{wang1998, linder2005}.

The results of this analysis are shown in Fig. \ref{fig:xi_ratio}, at three different redshifts, for all the models considered. The horizontal lines in each panel show the $\Lambda$CDM predictions computed with the linear biases by \citet{Tinker_bias_2010}, normalised at the $\sigma_8$ values of each model \citep{Marulli_2012A}. Thanks to the latter normalisation, all the considered models agree remarkably well with the $\Lambda$CDM predictions, particularly at scales beyond $10$\Mpch.
These results show that the effect of $f(R)$ gravity models, with or without massive neutrinos, on the redshift-space monopole of the halo 2PCF is strongly degenerate with $\sigma_8$, similarly to what was previously found in real space (see Fig. \ref{fig:bias_redshift}). This result also confirms what found by \citet{Marulli_2012A} for cDE models with massless neutrinos. Similar conclusions have been reached by \citet{Villaescusa_Navarro_2018}, who investigated the redshift-space clustering in massive neutrino cosmologies.

Due to the $\sigma_8$-degeneracy, the redshift-space 2PCF monopole alone is not sufficient to discriminate among these alternative cosmological frameworks. To break the degeneracy, the full 2D clustering information has to be extracted. As explained in Section \ref{sec:xiZS}, it is enough though to consider only the first two even multipoles, that is the monopole and the quadrupole (see Fig. \ref{fig:multipolesZS}).

To construct the likelihood, we consider the so-called {\em dispersion model} \citep[]{peacock1994}. Though it has been shown that it can introduce systematics in the linear growth rate measurements \citep[see e.g.][and references therein]{bianchi2012, delatorre2012, marulli2017}, the dispersion model is accurate enough for the purposes of the present work, that consists in quantifying the relative differences between $f(R)$ models and $\mathrm{\Lambda CDM}$. 

In the following, we briefly summarise the main equations of the dispersion model \citep[see e.g.][for more details]{delatorre2012}. Assuming the plane-parallel approximation, the redshift-space power spectrum of matter density fluctuations, $P^{\rm zs}(k,\mu)$, can be parametrised as follows:
\begin{equation}\label{eq:fisherPk}
P^{\rm zs}(k, \mu) =\left(1+\frac{f}{b} \mu^2 \right)^2 F(k, \mu, \Sigma_s) b^2 P(k, \mu)\,, 
\end{equation}
where the first term on the right-hand side is the linear Kaiser term, $b$ is the linear bias and $P(k,\mu)$ is the matter density power spectrum in real space. $F(k, \mu, \Sigma_s)$ is a damping function used to describe the FoG at small scales given by:
\begin{equation}\label{eq:streamming}
F(k, \mu, \Sigma_s)=\frac{1}{(1+k^2\mu^2 \Sigma_s^2)}\,,
\end{equation}
where the streaming scale $\Sigma_s$ is a free model parameter \citep{Kaiser_1987, Hamilton_1992, Fisher_1996}.
The model 2PCF multipoles are obtained by Fourier transforming the power spectrum multipoles, $P_l(k)$, as follows: 
\begin{equation}\label{eq:pkmultipoles}
P_l(k) = \frac{2l +1}{2}\int_{-1}^{+1} d\mu P(k,\mu) L_l(\mu)\,,
\end{equation} 
\begin{equation}\label{eq:ximultipolesmodel}
\xi_l(s) = \frac{i^l}{2\pi^2}\int dk k^2 P_l(k) j_l(ks)\,,
\end{equation} 
where $j_l$ are the $l^{th}$-order spherical Bessel functions \citep[for more details see e.g.][]{pezzotta2017}.

\begin{figure*}
	\includegraphics[width=\hsize]{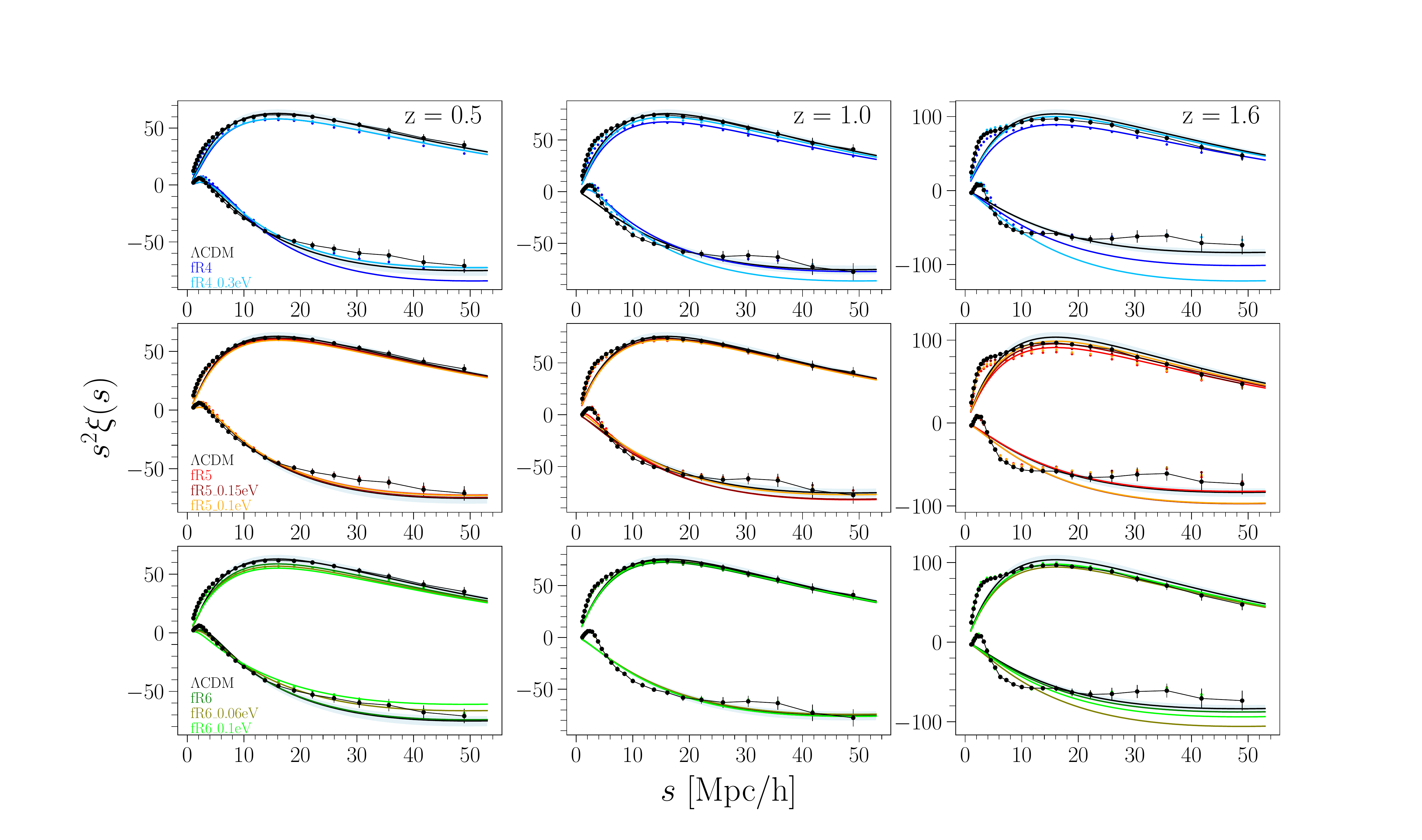}
	\caption{The measured redshift-space 2PCF multipole moments for the $\Lambda$CDM (points with errorbars) model compared to the best-fit posterior models, at three different redshifts ($z=0.5,~1$ and $1.6$, columns from left to right), for the different MG models (as labelled in the plot).
	\label{fig:2PCFmultipoles}}
\end{figure*}
\begin{figure*}
	\includegraphics[width=\hsize]{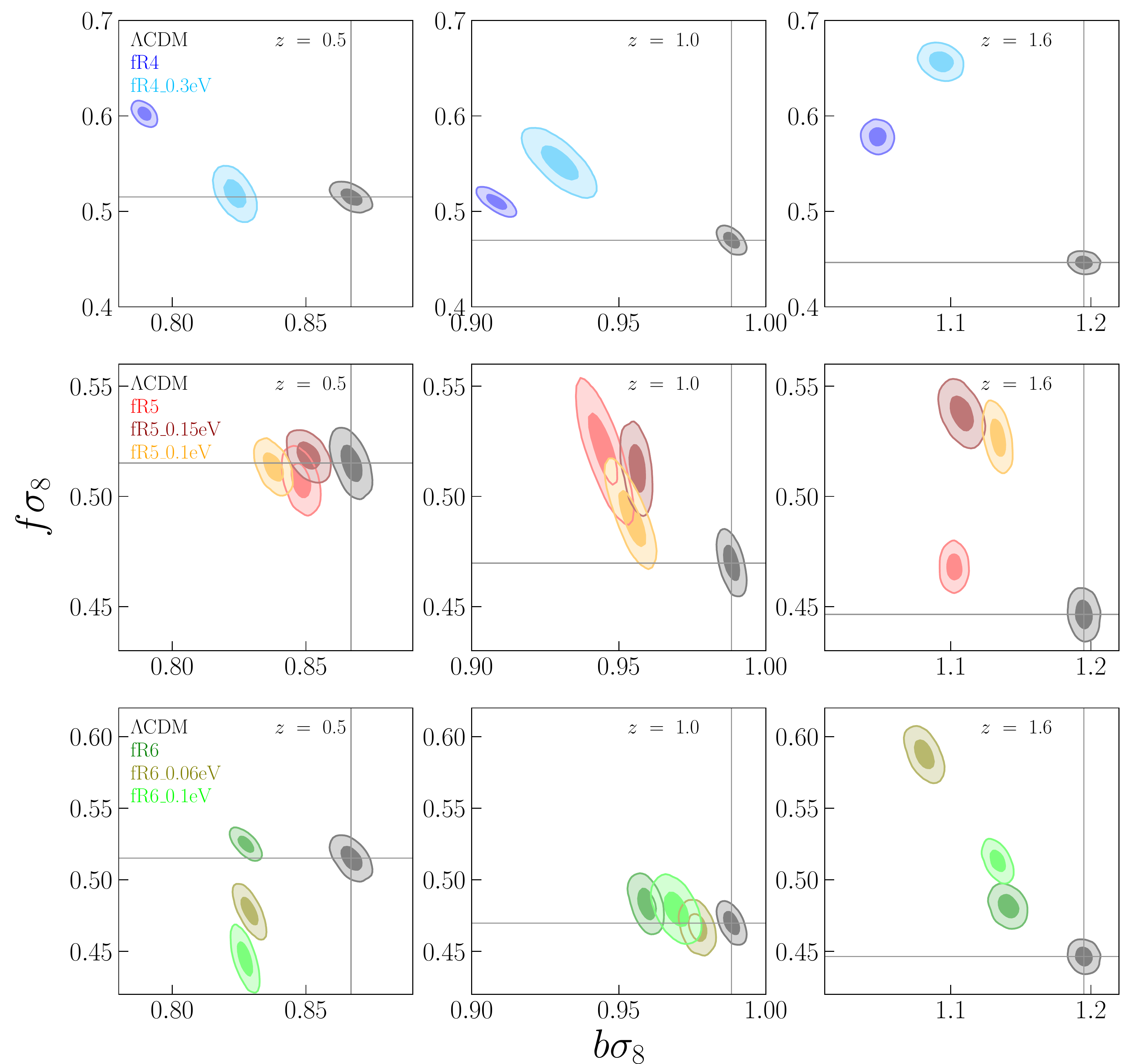}
	\caption{Posterior constraints at $1-2\sigma$ confidence levels in the $f\sigma_8-b\sigma_8$ plane, marginalised over $\Sigma_s$, obtained from the MCMC analysis of the redshift-space monopole and quadrupole moments of the 2PCF for the MG models are shown by different colours, as labelled. Panels in the columns, from left to right, refer to $z=0.5,~1$ and $1.6$.
	\label{fig:contourmodels}}
\end{figure*}

The dispersion model (Eqs. \ref{eq:fisherPk}, \ref{eq:streamming}) can be written in terms of three free parameters, $f\sigma_8$, $b\sigma_8$ and $\Sigma_S$, that we constrain by minimising numerically the negative log-likelihood:
\begin{equation}\label{eq:loglike}
-2\ln\mathcal{L} =\sum_{i,j=1}^N\left[\xi_l^D(s_i)-\xi_l^M(s_i)]C_l(s_i,s_j)^{-1}[\xi_l^D(s_j)-\xi_l^M(s_j)\right]\,,
\end{equation}
with $N$ being the number of bins at which the multipole moments are estimated, and the superscripts $D$ and $M$ referring to data and model, respectively.
The covariance matrix $C_l(s_i,s_j)$ is computed from the data with the bootstrap method:
\begin{equation}\label{eq:corrmatrix}
C_l(s_i,s_j)=\frac{1}{N_R-1}\sum_{n=1}^{N_R}\left[\xi_l^n(s_i)-\bar{\xi}_l(s_i)][\xi_l^n(s_j)-\bar{\xi}_l(s_j)\right]\,,
\end{equation}
where the indices $i$ and $j$ run over the 2PCF bins, $l = 0,~2$ correspond to the multipole moments considered, $\bar{\xi}_l = 1/N_R\sum_{n=1}^{N_R}\xi_l^n$ is the average multipole of the 2PCF, and $N_R=100$ is the number of realisations obtained by resampling the catalogues with the bootstrap method. To assess the posterior distributions of the three model parameters, a MCMC analysis is performed. The fitting analysis is limited to the scale range $10\leq r~[\mbox{Mpc}\,h^{-1}] \leq 50$, assuming flat priors in the ranges $0\leq f\sigma_8\leq 2$, $0\leq b\sigma_8\leq 3$ and $0\leq \Sigma_S\leq 2$.

\begin{figure*}
	\includegraphics[width=\hsize]{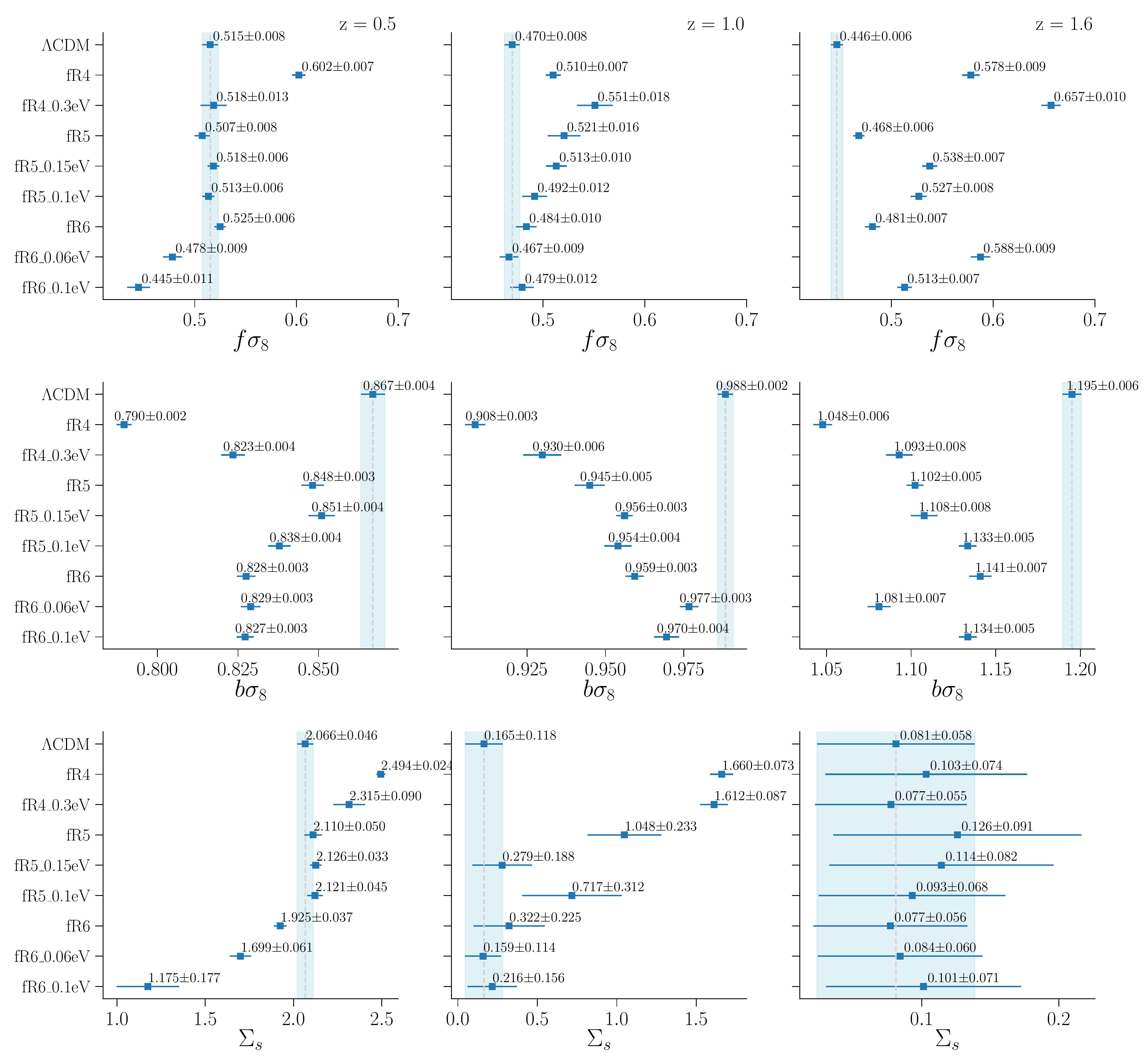}
	\caption{Posterior constraints at $1\sigma$ and $2\sigma$ confidence levels for $f\sigma_8$ (upper panels), $b\sigma_8$ (central panels) and $\Sigma_s$ (bottom panels), at three different redshifts $z=0.5, 1.0$ and $1.6$ from left to right, for all models considered in this work, obtained from the MCMC analysis of the redshift-space monopole and quadrupole moments of CDM haloes. The vertical shaded areas are centred on $\Lambda$CDM results, for comparison.
	\label{fig:comparisonparams}}
\end{figure*}

As an illustrative example, Fig. \ref{fig:breakingdegeneracy} shows the $f\sigma_8$-$b\sigma_8$ posterior constraints, marginalised over $\Sigma_s$, obtained from the MCMC analysis of $\xi_0$, $\xi_2$ and $\xi_0+\xi_2$ of a $\Lambda$CDM halo mock sample at $z=0.5$. As it is well known, a joint analysis of the redshift-space monopole and quadrupole is required to break the degeneracy between $f\sigma_8$ and $b\sigma_8$, as it is shown in the Figure.
We apply this analysis to all the \dustp mock catalogues. Fig. \ref{fig:2PCFmultipoles} shows the monopole and quadrupole measurements compared to best-fit model predictions. The latter are obtained by assuming the dispersion model, with $\Lambda$CDM power spectrum, normalised to the $\sigma_{8}$ values of each \dustp simulation. As in all previous plots, this method simulates the statistical analysis that would be performed if the real cosmological model of the Universe was one of the $f(R)$ assumed scenarios, with or without massive neutrinos, while a $\Lambda$CDM model was instead erroneously assumed to predict the DM clustering. 

Fig. \ref{fig:contourmodels} shows the $f\sigma_8$-$b\sigma_8$ posterior contours, at $1-2\sigma$, for all models and redshifts considered. This represents our main result: the alternative MG models considered in this work can be clearly discriminated at $z\gtrsim 1$, also in the presence of massive neutrinos whose masses are chosen to introduce strong degeneracies in linear real-space statistics. 

A final summary of all our $f\sigma_8$, $b\sigma_8$ and $\Sigma_s$ cosmological constraints is presented in Fig. \ref{fig:comparisonparams}. At low redshifts, the $f\sigma_8$ posteriors of almost all the $f(R)$ models considered appear statistically indistinguishable from $\Lambda$CDM (as already evident in Fig. \ref{fig:contourmodels}). Nevertheless, at higher redshifts they are clearly no more degenerate. This is an interesting result, given that the next-generation dark energy experiments, such as the ESA Euclid mission \citep{laureijs2011}, will mainly probe the high redshift ($z>1$) Universe. 
To investigate how our estimated uncertainties on $\Delta f\sigma_8$ and $\Delta b\sigma_8$ depend on the survey volume, we repeated our analysis on $5$ smaller sub-boxes, extracted from the original simulation snapshots, with increasing sides, $L_{\rm box}=350, 450, 550$ and $650$ Mpc$\,h^{-1}$. We found approximately linear relations between the estimated uncertainties and the survey volume. Considering the volume of surveys like Euclid, we expect that the uncertainties on both $f\sigma_8$ and $b\sigma_8$ will be about $10$ times smaller relative to the values estimated in the current analysis. However, there are many complications affecting the analysis on real data, that might significantly increase the estimated uncertainties. Reliable forecasts should include both statistical and systematic uncertainties possibly caused by observational effects, such as e.g. redshift measurement errors, photometric and spectrophotometric calibration, sky brightness variations, geometric selections.

\section{Conclusions}\label{sec:conclusions}
We investigated the clustering and redshift-space distortions of CDM haloes in MG models, with and without massive neutrinos. The present work is a follow-up to the analyses presented in \citet{Marulli_2012A}, who investigated the halo clustering properties in cDE cosmological scenarios. In particular, this paper extended the analysis to $f(R)$ models, investigating at the same time the effects of including massive neutrinos. Specific combinations of parameters in $f(R)$ gravity and neutrino masses are considered to investigate possible cosmic degeneracies in the spatial properties of the LSS of the Universe. The family of MG models analysed in this work mimics the $\Lambda$CDM background expansion on large scales, being also consistent with solar system constraints \citep{Hu_Sawicki_2007}. 

In this work we studied whether redshift-space distortions in the 2PCF multipole moments can be effective in breaking these cosmic degeneracies. The analysis has been performed using mock halo catalogues at different redshifts extracted from the \dustp runs, a set of N-body simulations of $f(R)$ models with and without massive neutrinos \citep{Giocoli_lensing_2018A, Peel_degeneracy_2018, Peel_2018_MG_MachiLearn, Merten_2018_dissection}. We considered intermediate scales, below $50$\Mpch, focusing on the first two even multipole moments of the 2PCF. We exploited a Bayesian statistical approach to assess posterior probability distributions for the three free parameters of the dispersion model $\{f\sigma_8,~b\sigma_8,~\Sigma_S\}$. The main result that came out from this analysis is that redshift-space distortions of 2PCF multipoles are effective probes to disentangle cosmic degeneracies, though only at large enough redshifts ($z\gtrsim 1$). In fact, the linear growth rate constraints obtained from all the analysed $f(R)$ mock catalogues are statistically distinguishable from $\Lambda$CDM predictions, at all redshifts but $z=0.5$, as shown in Fig. \ref{fig:contourmodels} and \ref{fig:comparisonparams}.

%%%%%%%%%%%%%%%%%%%%%%%%%%%%%%%%%%%%%%%%%%%%%%%%%%

\section*{Acknowledgements}
We acknowledge the grants ASI n.I/023/12/0, ASI-INAF n. 2018-23-HH.0 and PRIN MIUR 2015 ``Cosmology and Fundamental Physics: illuminating the Dark Universe with Euclid". The \dustp simulations discussed in this work have been performed and analysed on the Marconi supercomputing machine at Cineca thanks to the PRACE project SIMCODE1 (grant nr. 2016153604, P.I. M. Baldi) and on the computing facilities of the Computational Center for Particle and Astrophysics (C2PAP) and of the Leibniz Supercomputer Center (LRZ) under the project ID pr94ji. JEGF thanks financial support from ``Convocatoria Doctorados Nacionales 757 de COLCIENCIAS''.

%%%%%%%%%%%%%%%%%%%%%%%%%%%%%%%%%%%%%%%%%%%%%%%%%%
%%%%%%%%%%%%%%%%%%%% REFERENCES %%%%%%%%%%%%%%%%%%
\bibliographystyle{mnras}
\interlinepenalty=10000
\bibliography{MGNeutrinos.bib}
%%%%%%%%%%%%%%%%%%%%%%%%%%%%%%%%%%%%%%%%%%%%%%%%%%
% Don't change these lines
\bsp	% typesetting comment
\label{lastpage}
\end{document}